\documentclass[3p]{elsarticle}
\usepackage{amssymb}
\usepackage{multirow}
\usepackage{booktabs}
\usepackage{array}
\usepackage{graphicx}
\usepackage{graphics}
\usepackage{subfig}
\usepackage{threeparttable}

\usepackage{hyperref}

\newcommand{\AAA}{\mathcal{A}}

\begin{document}
\begin{frontmatter}
\title{Spam filtering by quantitative profiles}

\author[adresa2]{M. Grend\'ar}
\ead{marian.grendar@slovanet.net}
\author[adresa2]{J. \v Skutov\'a}
\ead{jana.skutova@slovanet.net}
\author[adresa2]{V. \v Spitalsk\'y}
\ead{vladimir.spitalsky@slovanet.net}

\address[adresa2]{Slovanet a.s., Z\'ahradn\'icka 151, 821 08 Bratislava, Slovakia}

\begin{abstract}
Instead of the ``bag-of-words'' representation, in the quantitative profile  approach to spam filtering and email categorization, an email is represented by an $m$-dimensional vector of numbers, with $m$ fixed in advance. Inspired by Sroufe et al. [Sroufe, P., Phithakkitnukoon, S.,   Dantu, R., and   Cangussu, J. (2010). Email shape analysis. In \emph{LNCS}, 5935, pp. 18-29] two instances of quantitative profiles are considered: line profile and character profile. Performance of these profiles is studied on the TREC 2007, CEAS 2008 and a private corpuses. At low computational costs, the two quantitative profiles achieve performance that is at least comparable to that of heuristic rules and naive Bayes.
\end{abstract}
\begin{keyword}
email categorization \sep spam filtering \sep quantitative profile \sep character profile  \sep line profile \sep Random Forest
\end{keyword}
\end{frontmatter}

\section{Introduction}

Spam is an unsolicited email message. From the receiver's perspective, spam is an annoyance and thus it is necessary to block its delivery, for instance, by filtering it out.

Traditional approach to spam filtering and email categorization that is based on heuristic rules,
naive Bayes filtering and/or text-mining suffers from several deficiencies. Among shortcomings of the traditional approach
there are high computational costs, language dependence, necessity to update the heuristic rules,  high number of rules, and vulnerability.

Instead of the ``bag-of-words'' representation, employed in text-mining and naive Bayes filtering, in the quantitative profile (\emph{QP}) approach that we propose, an email is represented by an $m$-dimensional vector of numbers with $m$ fixed in advance. Inspired by Sroufe, Phithakkitnukoon, Dantu, and Cangussu \cite{ESA}, two instances of \emph{QP}s are considered: line profile (\emph{LP}) and character profile (\emph{CP}). Informally put, the line profile of an email is a vector of lengths of the first $m$ lines. The character profile is a histogram of characters. Of course, many other \emph{QP}s are conceivable.

The main advantages of the two considered quantitative profiles are $i)$ sound performance, $ii)$ simple computability, $iii)$ language-independence, $iv)$ robustness to outlying emails, $v)$ high scalability and $vi)$ low vulnerability.

The considered two instances of the quantitative profile approach  perform comparably to the naive Bayes filtering and heuristics-based approaches and perform very well also in a multi-language, non-English communication. Furthermore, the satisfactory performance is attained by means of a small set of easily computable quantitative features. A performance study was done on the TREC 2007, CEAS 2008 and a private corpuses. To demonstrate the power of the considered \emph{QP}s, the profiles are obtained from raw emails, without any preprocessing. {Consequently, we intentionally ignore the structure of emails and character encoding}.

On the two \emph{QP}s, the Random Forest algorithm substantially outperforms other classification algorithms (SVM, LDA/QDA, logistic regression).
Thanks to the Random Forest, the \emph{QP} approach gains robustness to emails with extreme-valued profiles as well as high scalability.

In our view, classification and email categorization by  \emph{LP} (or \emph{CP}) should have low vulnerability.
For, the lines the lengths of which differentiates between spam and ham, change from corpus to corpus.
For instance, in the CEAS 2008 corpus, the most important for deciding between spam and ham are the lengths of the (10, 17, 15, 14, 16)-th lines, whilst in the TREC 2007 corpus they are the (5, 13, 6, 15, 7)-th lines. As the training corpus is usually not available to a spammer, it should be not easy to evade the \emph{LP} (or \emph{CP}) filter.

As a by-product of a performance study of \emph{CP} and \emph{LP}, we note that in the TREC 2007 and CEAS 2008 corpuses the number of header lines is capable of discriminating between spam and ham, at a rate that is, in our view, too high. Consequently, the corpuses lead to overly optimistic performance of spam filtering methods.

The paper is organized as follows. In the next section we formally introduce the \emph{QP}s mentioned above. Then, in Section~\ref{data}, we describe the three email corpuses used for assessment of the \emph{QP}s' performance. In Section~\ref{perf-measures} we describe measures used for performance evaluation. The results are summarized in Section~\ref{results}. In the concluding section some directions for future research are briefly discussed. All the computations were performed with R \cite{R}. To make the results reproducible, a supplementary  material including the source code was prepared, cf.~\cite{Rcode}.

\section{Quantitative profiles}
\label{QP}

The \emph{quantitative profile} (\emph{QP})  of an email is an $m$-dimensional vector of real numbers that represents the email. The dimension $m$ of the profile is set in advance, and it is the same for all emails. In this paper we consider two particular \emph{QP}s -- the line profile and the character profile.
These profiles can be introduced by means of a simple probabilistic model.

An email is represented as a realization of a vector random variable, that is generated by a hierarchical data generating process.
The length $n$ of an email is an integer-valued random variable, with the probability distribution $F_n$. Given the length,
the email is represented by a random vector $X_1^n = (X_1, \dots, X_n)$ from the probability distribution $F_{X_1^n\,|\,n}$ with  the support in $\AAA ^n$, where $\AAA = \{a_1, \dots, a_m\}$ is a finite set (alphabet) of size $m=|\AAA|$.

Then, the \emph{character profile} (\emph{CP}) of an email is an $m$-dimensional random vector
$\mathit{CP} = (\mathit{CP_1},\dots,\mathit{CP}_m)$, where
$$
\mathit{CP}_j = \sum_{i=1}^n I_{\{X_i = a_j\}},
\qquad j=1,\dots,m,
$$
and $I$ is the indicator function.

\begin{figure}[h!]
  \centering
  \subfloat[Email]{\label{email}
     \includegraphics[height=4cm]{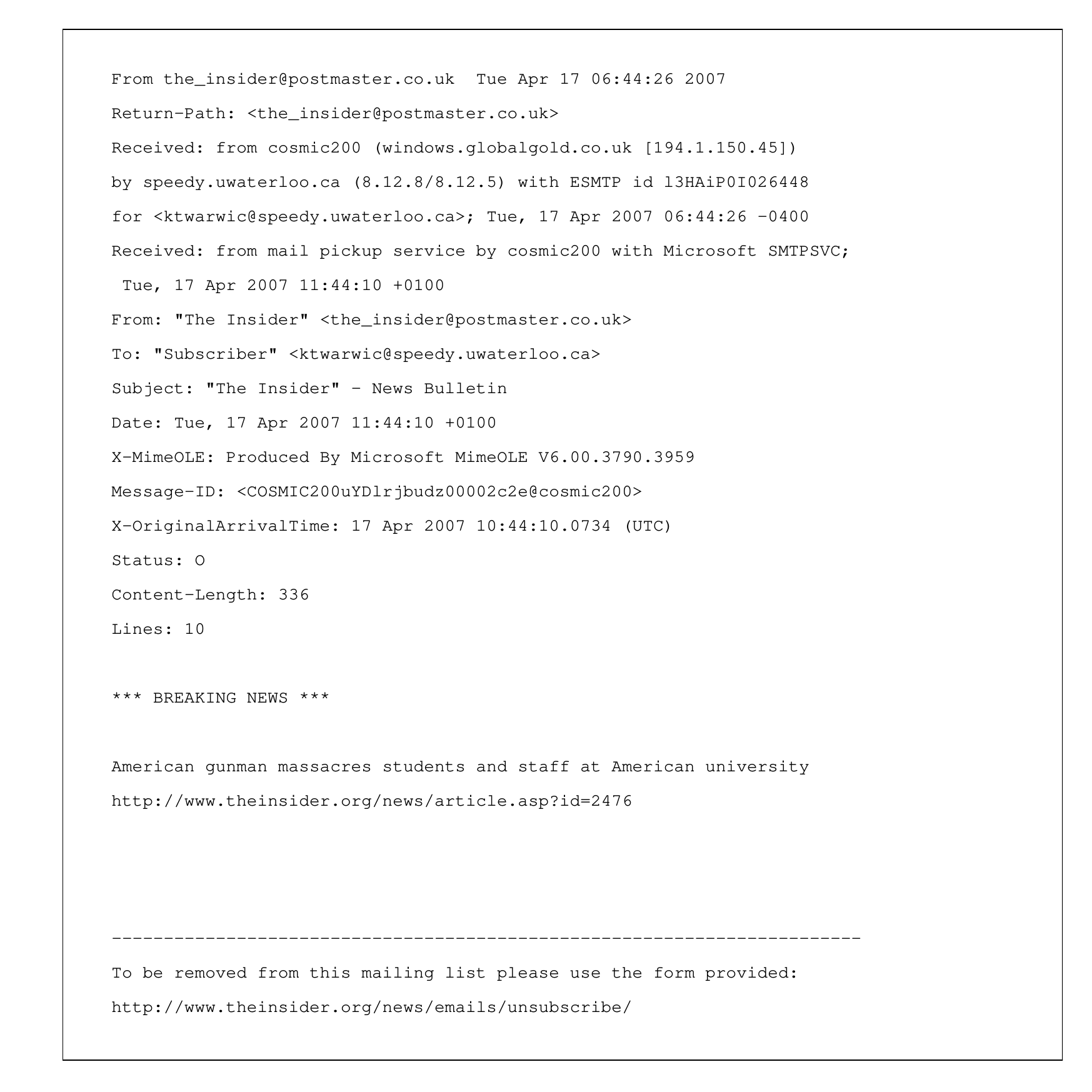}}
     \subfloat[Line profile]{\label{lp}
     \includegraphics[height=4cm]{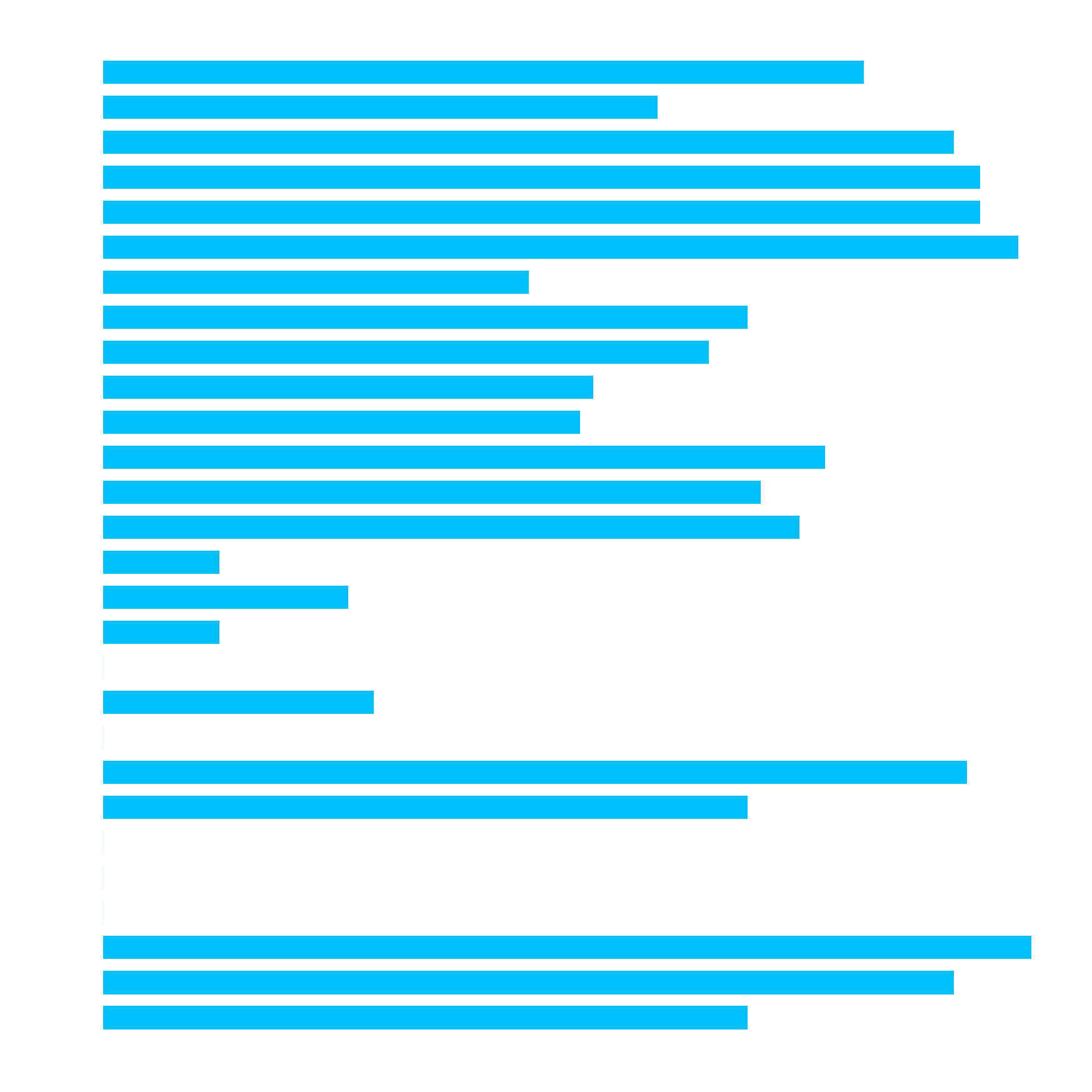}}
  \subfloat[Character profile]{\label{cp}
     \includegraphics[height=4cm]{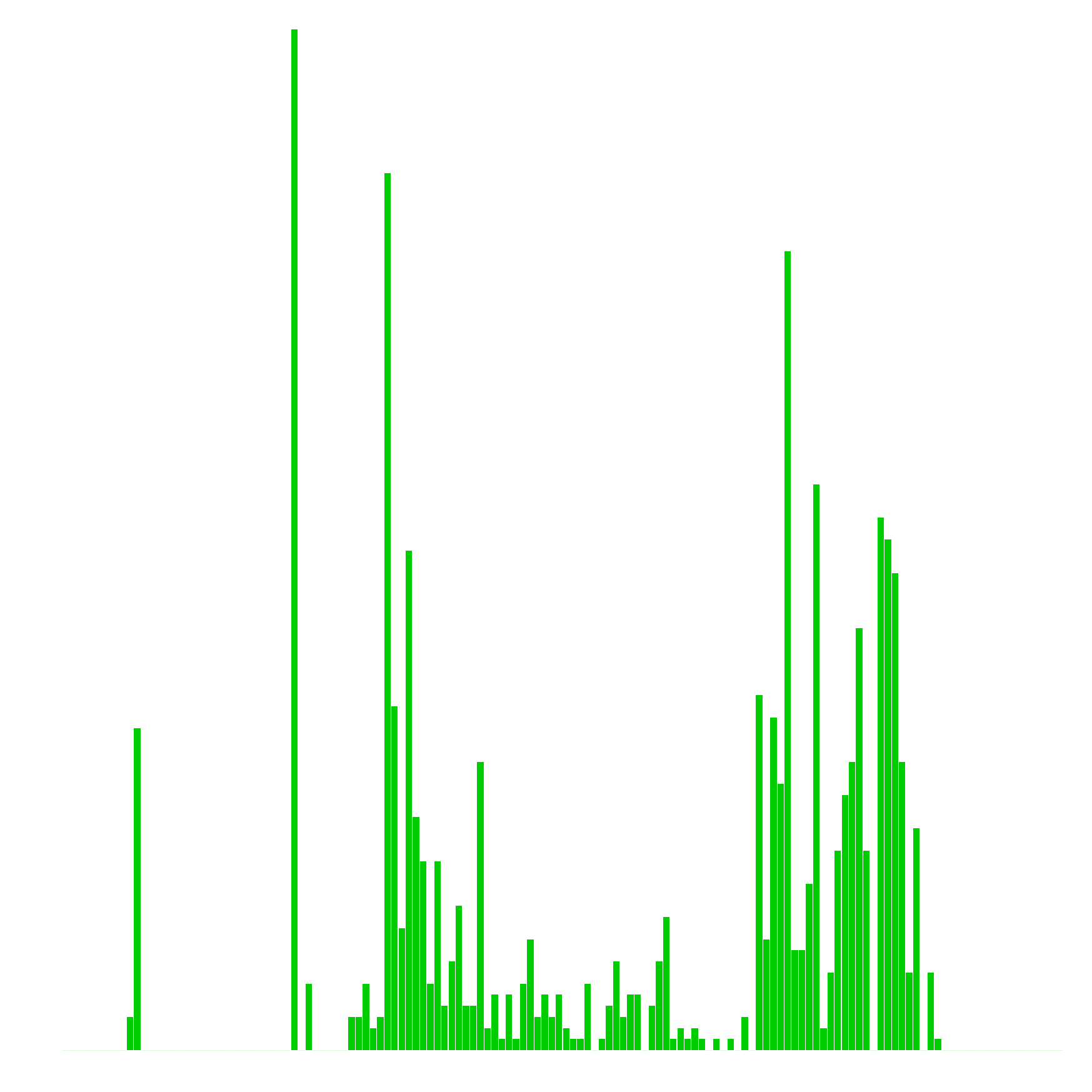}}
  \caption{Graphical representation of the line and character profiles of an email}
  \label{figure-emails}
\end{figure}

In order to introduce the other \emph{QP}, it is necessary to select a special character (or a subset of characters) from the alphabet.
Let $k$ be the number of occurrences of the special character in an email and let $\mathit{T}_j$ ($j=1,\dots,k$)
be the index of the $j$-th occurrence of the special character; put $T_0 = 0$. Then the \emph{binary profile} (\emph{BP}) of an email is defined as  a $\tilde k$-dimensional random vector
$\mathit{BP}=(\mathit{BP}_1,\dots,\mathit{BP}_{\tilde{k}})$, where
$$
 \mathit{BP}_j = \mathit{T}_j - \mathit{T}_{j-1} -1,
\qquad j=1,\dots,\tilde k.
$$
There, $\tilde k$ is the maximum allowable number of the occurrences of the special character and it is set in advance.
Hence, if $k > \tilde{k}$, the rest of the email is ignored. And if an email has $k<\tilde{k}$, $\mathit{BP}_j = 0$ for $j > k$.

In this work, an email is understood as a stream of bytes without any preprocessing, so that $\AAA$ is taken to be the ASCII character set.
The end-of-line is taken for the special character. Consequently, the binary profile becomes the \emph{line profile} (\emph{LP}) since
$BP_j$ is the length of  the $j$-th line of an email in bytes.
We fix the maximal number $\tilde{k}$ of considered lines  to $100$.

\section{Data sets}
\label{data}

To assess the performance of the two quantitative profiles, we consider three email corpuses: the publicly available TREC 2007 and CEAS 2008 corpuses, and a private corpus.

The TREC 2007 corpus \cite{TREC07} comprises over 75 000 emails (25~220 hams and 50~199 spams), of which approximately 67\% is spam, and the rest are ham emails. For the training phase we used the first 50~000 emails. The rest forms the test set. The ratio of spam to ham in the training and test sets is approximately 2:1.

The other publicly available corpus, used for the performance analysis of \emph{QP}s, is CEAS 2008
\cite{CEAS}. It consists of 137~705 emails (27~126 hams and 110~579 spams).
The corpus was hand labeled \cite{CEASpopis}. To form the training set, we used the first 90~000 emails. The test set comprises the remaining emails. The ratio of spam to ham in the training and test sets is approximately 4:1.

Performance of spam filtering algorithms is typically assessed on English language corpuses, such as the above mentioned TREC and CEAS.
When applied to non-English emails, their performance may be different. Due to the support from a Slovak internet services provider, we enjoyed the opportunity of access to a private corpus created in 2010, that comprises mainly non-English emails. Structure of the corpus is summarized in
Table~\ref{tab1}. The training set we used consists of 11~050 emails and the test set consists of 12~200 emails.

\begin{table}[h!]
\caption{{Composition of the private corpus}}
\begin{center}

\begin{tabular}{!{\vrule width 1pt}c!{\vrule width 1pt}c|c|c|c!{\vrule width 1pt}c!{\vrule width 1pt}}
  \noalign{\hrule height 1pt}
  corpus & s.ham & advert & notify & spam & total\\
  \noalign{\hrule height 1pt}
  train & 6837 & 1611 & 1225 & 1377 & 11~050\\
  \hline
  test & 3650 & 1409 & 5758 & 1383 & 12~200\\
  \noalign{\hrule height 1pt}
\end{tabular}
\end{center}
\label{tab1}
\end{table}

The private corpus was hand labeled. Emails were placed into one of the two groups: ham and spam. Moreover, ham was divided into advert, solicited ham (denoted s.ham) and notify.
Based on the language of the major part of an email, the emails were placed into one of the four language groups: Slovak/Czech, English, German, and other. The corpus comprises 64\% of the Slovak/Czech ham in the training set and 38\% in the test set.

\section{Classification algorithm and performance measures}
\label{perf-measures}

Quantitative profiles serve as an input to a classification algorithm. In this work we use the Random Forest classifier, introduced by Breiman \cite{RF} and ported to R by Liaw and Wiener \cite{Liaw}, with the default settings. We have employed also LDA, logistic regression, LASSO and SVM \cite{Friedman}, but these methods performed much worse.

To evaluate \emph{QPs'} performance, we calculate the false positive rate $\mathit{fpr} = FP/(TN + FP)$ and the false negative rate
$\mathit{fnr} = FN/(TP + FN)$, where
\emph{TP} (\emph{TN}) stands for the number of true positive (true negative) emails, i.e.~the correctly recognized spam (ham) emails; and \emph{FP} (\emph{FN}) stands for the number of false positive (false negative) emails, i.e.~the incorrectly recognized ham (spam) emails, respectively.
We also present the receiver operating characteristic (\emph{ROC}) curve, i.e.~the graph of the true positive rate  vs.~the false positive rate, obtained as functions of the decision threshold. The area \emph{AUC} under the \emph{ROC} curve is also reported.

\section{Results}
\label{results}

In this section we summarize performance of the basic \emph{QP}s on the three corpuses mentioned above.

\subsection{Comparison of quantitative profiles with SpamAssassin and Bogofilter}

For the sake of comparison, we report also results for
SpamAssassin (\emph{SA}) \cite{SA}, version 3.3.1, off-line and without the Bayes filter, and Bogofilter (\emph{BF}) \cite{Bogofilter},
version 1.2.2 with the default configuration. On the private corpus, the output from \emph{SA} was processed by the Random Forest classification algorithm, as it attains much better performance than \emph{SA} with the default weights. In addition to much better performance, it allows for email categorization, which is impossible with the default \emph{SA}. On the public corpuses, however, the default \emph{SA} classification performs better. Bogofilter allows a binary classification only. \emph{BF} was learnt in the batch mode.

\begin{table}[h!]
 \caption{\emph{fnr} (\%) at fixed $\mathit{fpr} = 0.5\%$ or $\mathit{fpr} = 1\%$}
\begin{center}
\begin{tabular}{!{\vrule width 1pt}l!{\vrule width 1pt}rr!{\vrule width 1pt}rr!{\vrule width 1pt}rr!{\vrule width 1pt}}
  \noalign{\hrule height 1pt}
   & \multicolumn{2}{c!{\vrule width 1pt}}{private}
   & \multicolumn{2}{c!{\vrule width 1pt}}{TREC07}
   & \multicolumn{2}{c!{\vrule width 1pt}}{CEAS08}  \\
  \cline{2-7}
   filter & at 0.5\% & at 1\%   & at 0.5\% & at 1\% & at 0.5\% & at 1\%\\
  \noalign{\hrule height 1pt}
  \emph{CP} & 14.39 & 11.64 &  2.53 & 0.49 & 4.38 & 4.25 \\
  \hline
  \emph{LP} & 21.33 & 20.10 & 0.30 & 0.13 & 0.39 & 0.27 \\
  \noalign{\hrule height 1pt}
  \emph{SA} & 12.68 & 10.10 & 35.87 & 30.51 & 76.14 & 69.92 \\
  \hline
  \emph{BF} & 13.05 & 7.38 & 0.40 & 0.06 & 0.47 & 0.36 \\
 \noalign{\hrule height 1pt}
\end{tabular}
\end{center}
\label{tab2}
\end{table}

\begin{figure}[h!]
  \centering
  \subfloat[private corpus]{\label{roc-private}
     \includegraphics[height=5.25cm]{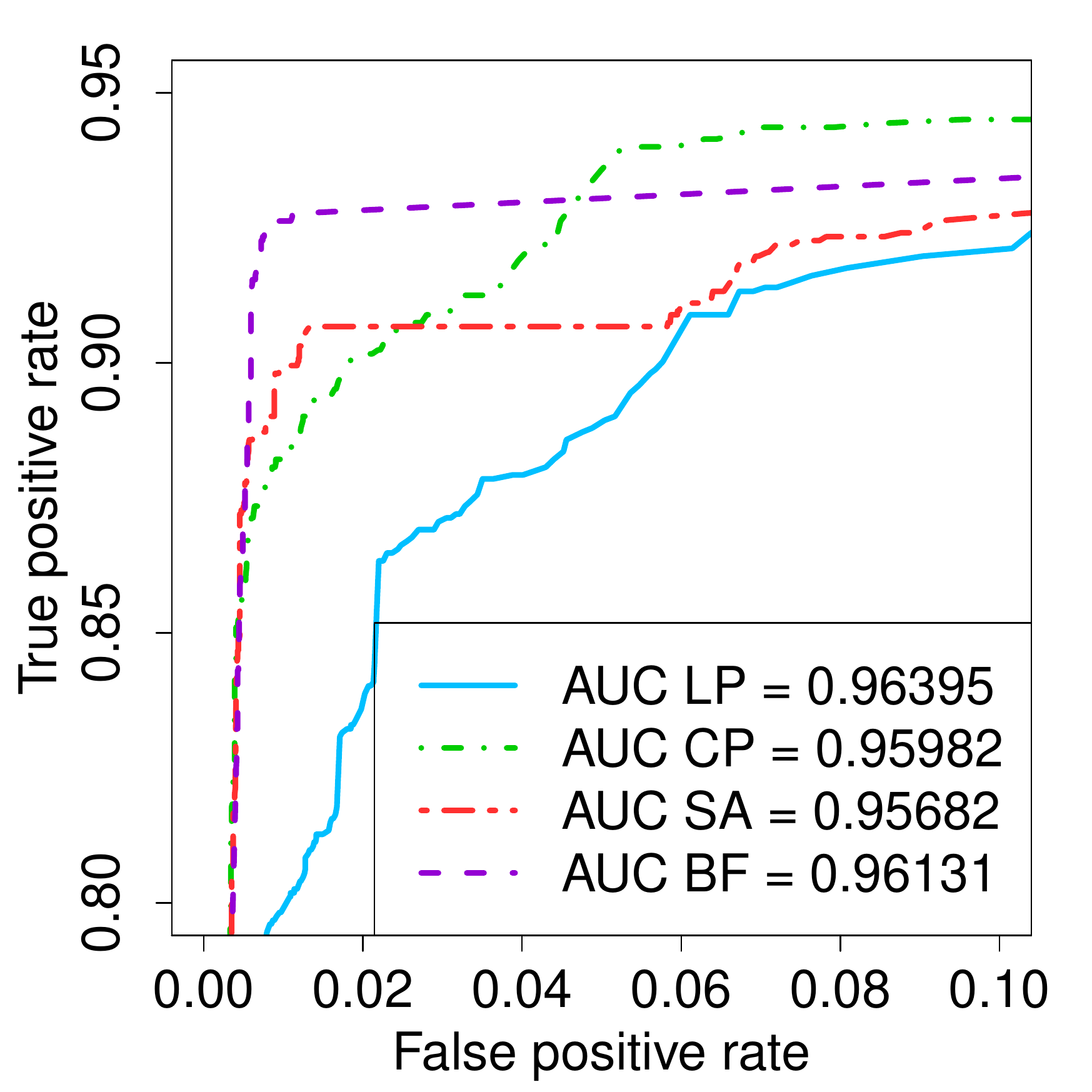}}
  \subfloat[TREC 2007 corpus]{\label{roc-trec}
     \includegraphics[height=5.25cm]{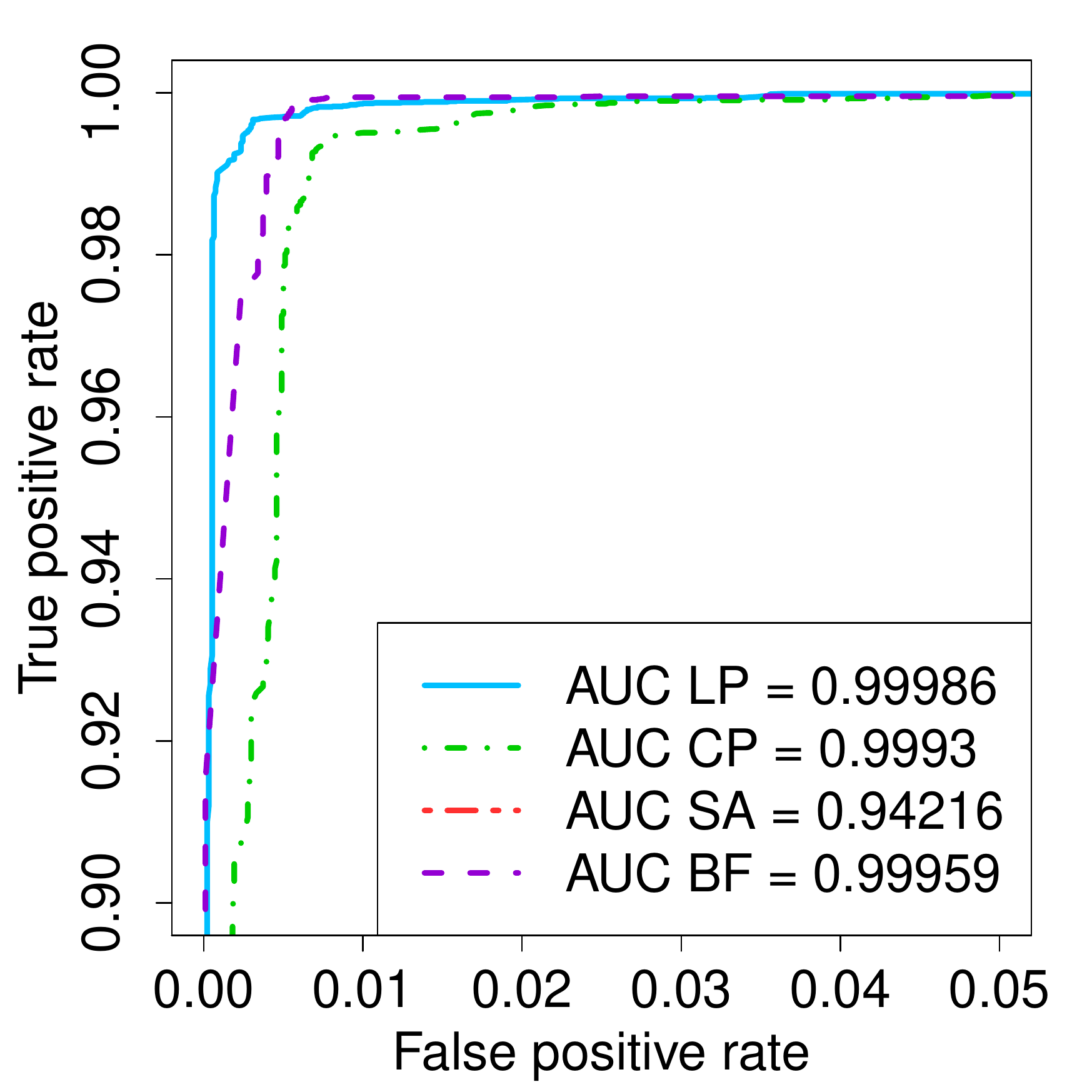}}
  \subfloat[CEAS 2008 corpus]{\label{roc-ceas}
     \includegraphics[height=5.25cm]{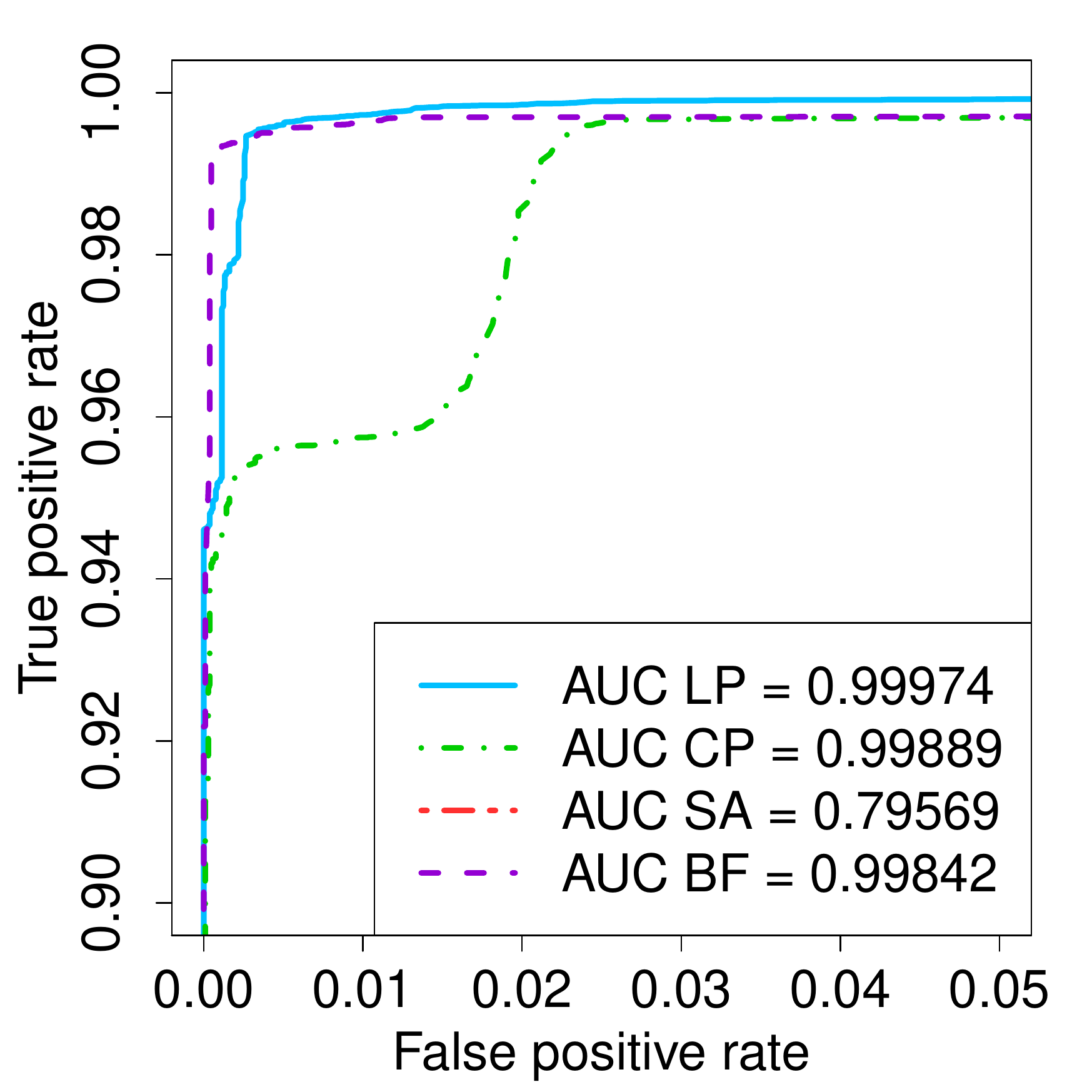}}
  \caption{\emph{ROC} curves}
  \label{ROC curves}
\end{figure}

On the private corpus, for $\mathit{fpr}=0.5\%$ the best performance is attained by \emph{SA} and \emph{BF}, and \emph{CP} performs only slightly worse, cf.~Table~\ref{tab2} and Figure~\ref{ROC curves}. \emph{LP} is slightly less effective, and it attains \emph{fnr} around $21\%$ at $\mathit{fpr}=0.5\%$.

With the exception of \emph{SA}, on the public corpuses all the studied filters  attain much better performance than on the private corpus, see also Section~\ref{What}. The two \emph{QP}s perform much better than \emph{SA} as Table~2 as well as Figures~\ref{roc-trec} and \ref{roc-ceas} indicate. The line profile attains better performance (smaller \emph{fnr}) than \emph{BF} at the $0.5\%$ level of \emph{fpr}.

\subsection{Email categorization for the private corpus}

On the private corpus, performance of \emph{CP} and especially \emph{LP} in spam filtering (i.e.~binary categorization) is worse than that of \emph{SA} and \emph{BF}. However, in categorization of emails  into one of the four categories, both \emph{CP} and \emph{LP} perform much better in the most interesting category of solicited ham. Misclassification table for \emph{CP} and \emph{SA} is in  Table~\ref{tab77}.

\begin{table}[h]
 \caption{\emph{CP} and \emph{SA} confusion tables for categorization, private corpus}
\begin{center}
\begin{tabular}{!{\vrule width 1pt}l!{\vrule width 1pt}r|r|r|r
  !{\vrule width 1pt}r|r|r|r!{\vrule width 1pt}}
\noalign{\hrule height 1pt}
  & \multicolumn{4}{c!{\vrule width 1pt}}{\emph{CP}}
  & \multicolumn{4}{c!{\vrule width 1pt}}{\emph{SA}}
\\
  \cline{2-9}
 count & advert & s.ham & notify & spam & advert & s.ham & notify & spam
 \\\noalign{\hrule height 1pt}
advert & 530 & 837 & 10 & 32 & 515 & 830 & 17 & 47
 \\\hline
ham & 26 & 3597 & 27 & 0 & 95 & 3290 & 248 & 17
 \\\hline
notify & 20 & 237 & 5499 & 2 & 24 & 235 & 5498 & 1
 \\\hline
spam & 37 & 221 & 12 & 1113 & 44 & 161 & 14 & 1164
 \\\noalign{\hrule height 1pt}
\end{tabular}
\end{center}
\label{tab77}
\end{table}

\subsection{Comparison with email shape analysis}

Sroufe et al.~\cite{ESA} suggest to filter spam by means of its shape, which the authors define (using our terminology) as a smoothed line profile of email body, where smoothing is performed by the kernel smoother. Sroufe et al. also report the total error of $30\%$, based on a preliminary study on the TREC corpus. Further, the authors find the performance very good, 'considering that no content or context was even referenced', cf.~\cite{ESA}, p.~26.   The line profile, that is inspired by the email shape analysis, attains on the TREC corpus $\mathit{fpr} = 4.23\%$ a $\mathit{fnr} = 17.00\%$, when the threshold is not optimized and email headers are intentionally not taken into account, cf. Section~\ref{What}. This gives the total error around $12.29\%$ and indicates  that smoothing is unnecessary.

\subsection{Reduction of the feature space}

It is also important to know to what extent the dimension of the \emph{QP} feature space can be reduced without substantive reduction of the classifier's performance. To this end  the top 20 and the top 50 features were considered, where the ranking of features was provided by the Random Forest's measure of the mean decrease of accuracy. The study was done on the private corpus and solely the email body was considered.

In the binary classification the top 20 features of the line profile attain essentially the same performance as the entire line profile of the length 100.
In the case of \emph{CP}, to attain the full-set performance, the top 50 features out of 256 are needed. The same holds for \emph{SA}. However, in the case of the email categorization it is not possible to reduce the dimension of \emph{SA} features without substantive decrease in accuracy.

\subsection{Why are TREC and CEAS misleading?}
\label{What}

\begin{figure}[h!]
  \centering
     \includegraphics[height=5cm]{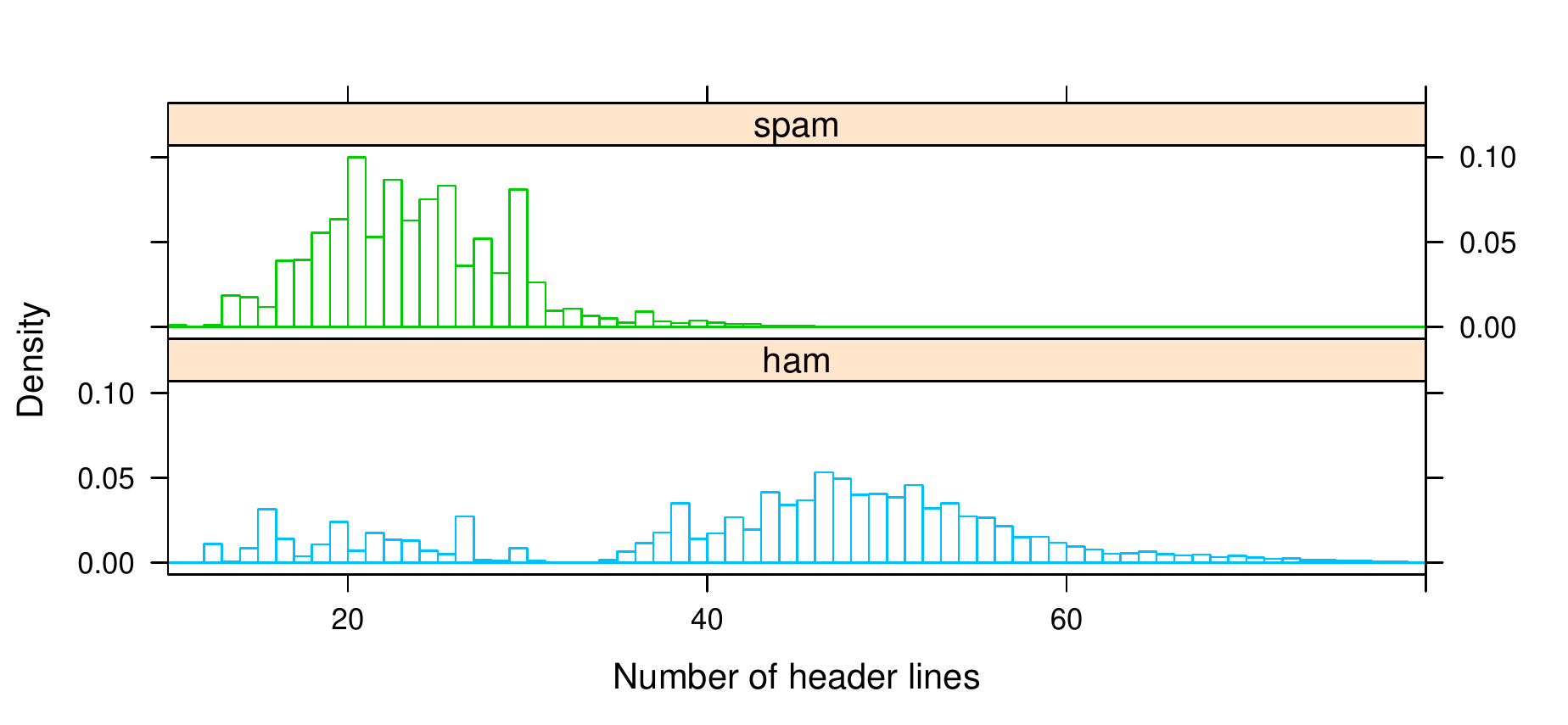}
  \caption{Distribution of emails with respect to the number of header lines, for spam and ham, in the TREC 2007 corpus}
  \label{figure-red-blue}
\end{figure}

All the considered email filters except of \emph{SA} attain much better performance on the public corpuses than on the private one; cf.~Table~2. In search for explanation we have noted that in the public corpuses, unlike to the private corpus, the number of header lines contains information that is substantive for spam filtering. Figure~\ref{figure-red-blue} depicts the distribution of the number of lines in header, for spam and ham, in the TREC corpus.

Once the email header is not taken into account, and solely the email body is processed, performance of \emph{LP} and \emph{CP} worsens and becomes comparable to that on the private corpus; cf.~Table~\ref{tab:HeaderBody}.  In Table~\ref{tab:HeaderBody}, $\emph{LPH}$ ($\emph{CPH}$) denotes the line (character) profile of email header and $\emph{LPB}$ ($\emph{CPB}$) denotes the line (character) profile of email body, respectively.

\begin{table}[h!]
\caption{\emph{fnr} (\%) at fixed $\mathit{fpr} = 0.5\%$ or $\mathit{fpr} = 1\%$}
\begin{center}
\begin{tabular}{!{\vrule width 1pt}l!{\vrule width 1pt}rr!{\vrule width 1pt}rr!{\vrule width 1pt}rr!{\vrule width 1pt}}
  \noalign{\hrule height 1pt}
   & \multicolumn{2}{c!{\vrule width 1pt}}{private}
   & \multicolumn{2}{c!{\vrule width 1pt}}{TREC07}
   & \multicolumn{2}{c!{\vrule width 1pt}}{CEAS08}
  \\
  \cline{2-7}
   filter & at 0.5\% & at 1\%   & at 0.5\% & at 1\% & at 0.5\% & at 1\%\\
  \noalign{\hrule height 1pt}
  \emph{CPH} & 16.51 & 13.71 & 0.19 & 0.05 & 0.78 & 0.30 \\
  \hline
  \emph{LPH} & 18.06 & 15.11 & 1.78 & 0.12 & 0.68 & 0.36 \\
  \noalign{\hrule height 1pt}
  \emph{CPB} & 14.24 & 11.92 & 15.05 & 5.47 & 4.70 & 4.51 \\
  \hline
  \emph{LPB} & 21.26 & 18.58 & 45.01 &  43.67 & 7.07 & 6.35 \\
  \noalign{\hrule height 1pt}
\end{tabular}
\end{center}
\label{tab:HeaderBody}
\end{table}

The decline of performance of \emph{CP} and \emph{LP} caused by exclusion of email headers supports the hypo\-thesis that in the TREC 2007 and CEAS 2008 corpuses the profiles of headers carry a substantive information for discriminating between spam and ham.

\subsection{Summary of the performance study}

The empirical study implies that the simple and easily obtainable line and character profiles attain at least comparable performance as the optimally tuned SpamAssassin, which is based on hundreds of fixed rules, and the performance of character profiles is close to that of Bogofilter. Particularly, on the public corpuses \emph{LP} is better than \emph{BF} and \emph{SA}. On the private corpus \emph{CP} attains comparable performance as \emph{BF} and \emph{SA}, and \emph{LP} is slightly worse.

\section{Conclusions}

Motivated by Sroufe et al. \cite{ESA}, we have proposed the quantitative profile  approach to email classification. In this report we explored two quantitative profiles, the line profile and the character profile. The profiles are obtained from raw emails, without any preprocessing. The computational costs of the two profiles are minimal. Performance of the profiles was studied on the TREC 2007, CEAS 2008 corpuses and a private, multi-lingual corpus. The two quantitative profiles attained at least comparable performance as the optimally tuned SpamAssassin and the batch-mode learnt Bogofilter. Besides the good performance, the two quantitative profiles  are language independent and the resulting filter is robust to outlying emails, highly scalable and has low vulnerability.

As a by-product, we have noted that the number of header lines in the TREC 2007 and CEAS 2008 corpuses contain rather strong information on the email class. The corpuses thus lead to overly optimistic performance of spam filters.

In the near future we plan to explore quantitative profiles based on size and structure of emails, on the symbolic dynamics, and another instances of the binary profile. Also, the profiles are worth employing in a semi-supervised email categorization.

\section{Acknowledgement}

Stimulating feedback from J{\' a}n Gallo and Stanislav Z{\' a}ri{\v s} is gratefully acknowledged. This paper was prepared as a part of the project ``SPAMIA'',  M\v S SR – 3709/2010-11, supported by the Ministry of Education, Science, Research and Sport of the Slovak Republic, under the heading of the state budget support for research and development.

\end{document}